\documentclass[twocolumn,showpacs,superscriptaddress,amsmath,amssymb]{revtex4}

\usepackage{graphicx}%
\usepackage{dcolumn}%

\begin{document}

\title{Life of the nodal quasiparticles in Bi-2212 as seen by ARPES}

\author{A. A. Kordyuk}
\affiliation{Institute for Solid State Research, IFW-Dresden, Helmholtzstr.~20, D-01069 Dresden, Germany}
\affiliation{Institute of Metal Physics of National Academy of Sciences of Ukraine, 03142 Kyiv, Ukraine}

\author{S. V. Borisenko}
\author{A. Koitzsch}
\author{J. Fink}
\author{M. Knupfer}
\author{B. B\"uchner}
\affiliation{Institute for Solid State Research, IFW-Dresden, Helmholtzstr.~20, D-01069 Dresden, Germany}

\author{H. Berger}
\affiliation{Institute of Physics of Complex Matter, EPFL, CH-1015 Lausanne, Switzerland}

\date{April 30, 2004}%

\begin{abstract}
While the pronounced doping dependence of the quasiparticle spectral weight in the antinodal region of the superconducting cuprates, as seen by ARPES, unambiguously points to the magnetic origin of the strong electron-boson coupling there, the nature of the electron scattering in the nodal direction remained unclear. Here we present a short review of our recent detailed investigations of the nodal direction of Bi-2212. Our findings prove the existence of well defined quasiparticles even in the pseudogap state and show that the essential part of the quasiparticle scattering rate, which appears on top of Auger-like electron-electron interaction, also implies a magnetic origin.
\end{abstract}


\maketitle

% main text

\section{ARPES view}

Angle-resolved photoemission spectroscopy (ARPES) \cite{ARPESreviews} provides a direct view on the density of low energy electronic excited states in solids---the 2D detector of the electron analysers used in modern ARPES is just a window into momentum-energy space of 2D compounds. A snapshot through this window stores the quasiparticle spectral weight in the momentum-energy co-ordinates \cite{VallaSci99,BogdanovPRL00,KaminskiPRL01,BorisenkoPRB2001}. Being essentially two-dimensional, the superconducting cuprates are a perfect example of the "arpesable" compounds \cite{ARPESreviews}. All the interactions of the electrons which are responsible for their unusual normal and superconducting properties are encapsulated in such snapshots, and success in understanding of the nature of electronic interactions in the cuprates depends, in the first place, on how clear the ARPES window is. Then, the experimental experience (namely, how many different snapshots have been taken and made out) comes into play. But taking into account a number of parameters (e.g. temperature and doping) which cause a redistribution of the quasiparticle spectral weight, the detailed exploration of the momentum-energy space, even for one compound, will take ages of experimental work. 

\begin{figure}[!b]
\includegraphics[width=7.4cm]{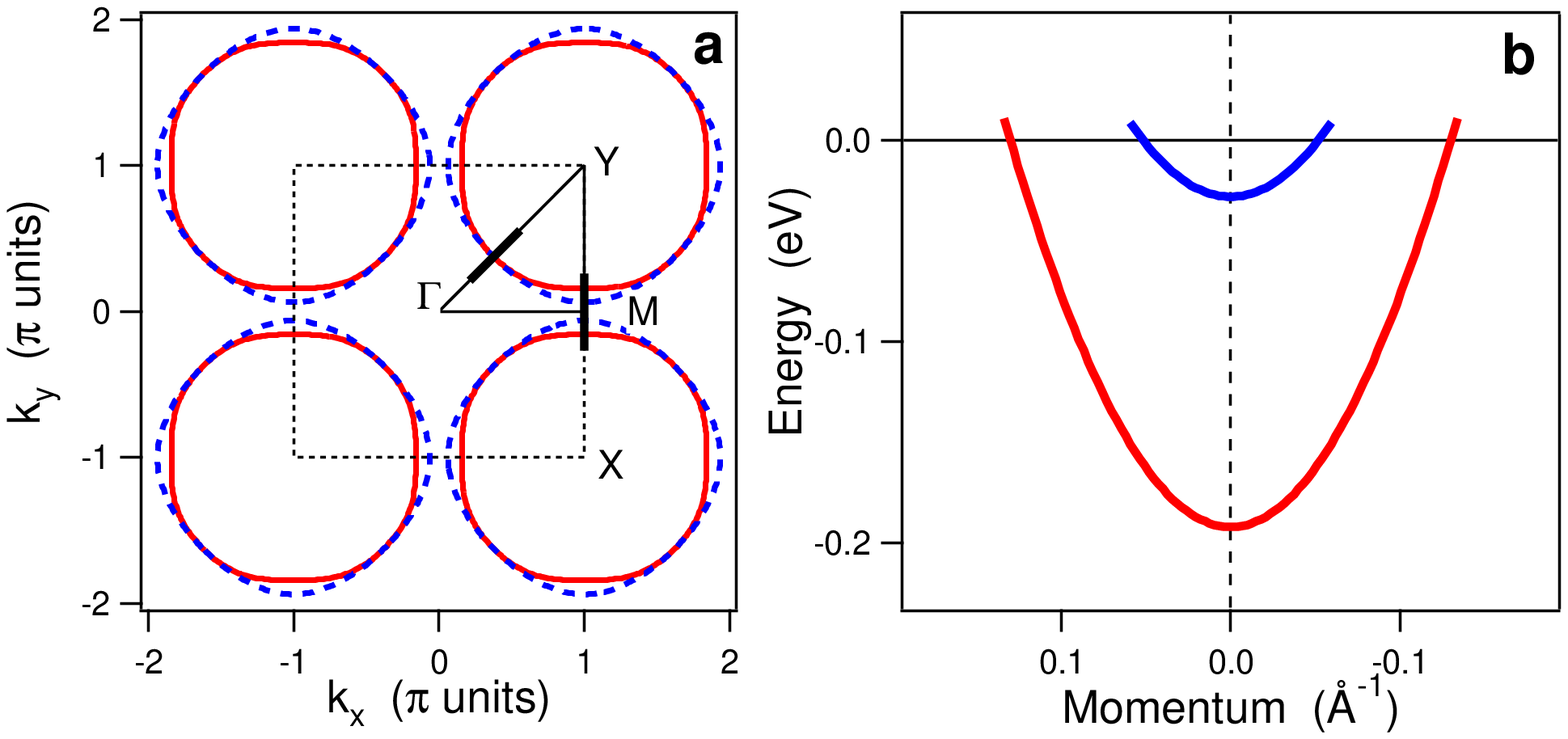} \\%
\includegraphics[width=7.6cm]{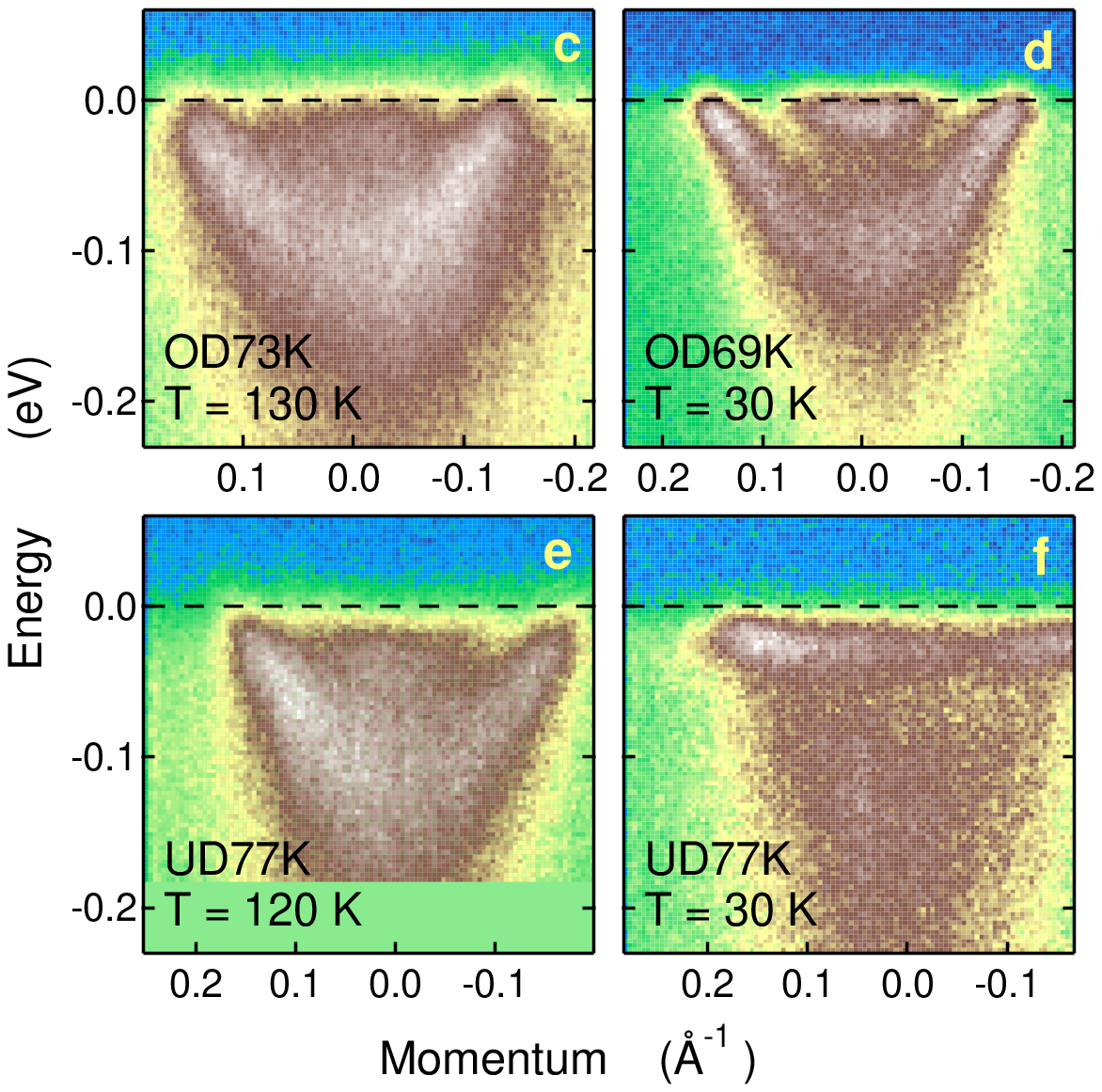}%
\caption{\label{XMY} Electronic band structure of an overdoped Bi-2212 \cite{KordyukPRB2003}: Fermi surfaces (a) and the "XMY" cut (b). ARPES snapshots taken along the XMY direction of the BZ \cite{BorisenkoPRL2003} for the overdoped (c, d) and underdoped (e, f) samples above (c, e) and below (d, f) $T_c$.}
\end{figure}

Leaving such a global task for the nearest future, one can focus on the two cuts in the Brillouin zone (BZ): nodal and antinodal directions (see Fig.~\ref{XMY} a). These regions represent an inherent anisotropy of the electronic interactions in the cuprates which appear in anisotropy of the superconducting gap \cite{SG}, pseudo-gap \cite{PG}, and coupling strength \cite{KimPRL2003} (or scattering in general). While the pronounced doping dependence of the quasiparticle spectral weight in the antinodal region of the BZ unambiguously points out to the magnetic origin of the strong electron-boson coupling seen by ARPES \cite{KimPRL2003,BorisenkoPRL2003,GromkoPRB2003} (see Fig.~\ref{XMY} c-f), the nature of the electron scattering in the nodal direction remaines unclear. Here we report the results of a detailed investigation of the nodal direction of Bi-2212 in a wide range of doping, temperature and excitation energy. We have found that although the electronic band structure along the nodal direction remains complex due to non-vanishing bilayer splitting, the quasiparticle spectral weight distribution from each split band can be self-consistently described within the quasiparticle self-energy approach. The scattering rate, on the other hand, can be considered as a sum of two main channels: the doping independent channel can be well understood in terms of the conventional Fermi liquid model, while the additional doping dependent channel implies a magnetic origin. 

\section{Experimental cornerstones}

The experimental details can be found elsewhere \cite{BorisenkoPRB2001,KimPRL2003,BorisenkoPRL2003}, but here we highlight the cornerstones which are peculiar to our experiments. They are: the precise cryo manipulator, the wide excitation energy range, and the superstructure-free samples.

The precise cryo manipulator operates in the controlled temperature range from 20 K to 400 K and allows us to translate the sample in three perpendicular directions and rotate it around three perpendicular axes in steps of 0.1$^\circ$, that secures the precise positioning and easy motion of the ARPES window in the momentum space.

The photons of different energy and polarisation have appeared to be an extremely useful tool to selectively excite the electrons from different bands \cite{KordyukPRL2002,BorisenkoPRB2004}. As a light source we use a He discharge lamp, linearly polarized light from a high resolution beamline (U125/1-PGM at BESSY) with a wide excitation energy range (17--600 eV), or circularly polarized light (4.2R beamline "Circular Polarization" at ELETTRA).

The well known problem for ARPES on Bi-2212 is a "5x1" superstructure which produces a number of diffraction replicas along the BZ diagonal ($\Gamma$Y direction) \cite{DingPRL1996,BorisenkoPRL2000}. This highly complicates the analysis of the spectra taken from certain areas of the BZ (e.g., the antinodal region \cite{KordyukPRL2002,BorisenkoCPandSS}) or, if the main features and replicas are spatially separated (e.g. along $\Gamma$Y direction), reduces the photocurrent intensity. Another important point about Bi-2212 samples, which becomes crucial now \cite{Timusk,Shen}, is that one needs to know precisely their doping level. We studied both the led-doped superstructure-free Bi(Pb)-2212 and reach of replicas pure Bi-2212. However, as the main line, we use the well characterized superstructure-free samples of a wide doping range ($x =$ 0.11 -- 0.22), for which the charge carrier densities have been derived from the measured Fermi-surface area (and appeared to be consistent with their $T_c$) \cite{KordyukPRB2002} and the tight-binding parameters have been determined \cite{KordyukPRB2003}. The parameters of the band structure of the pure Bi-2212 samples which we have measured are also in agreement with their doping level estimated from the $T_c$ measurements.
 
\section{Complex structure}

\begin{figure}[t]
\includegraphics[width=4.25cm]{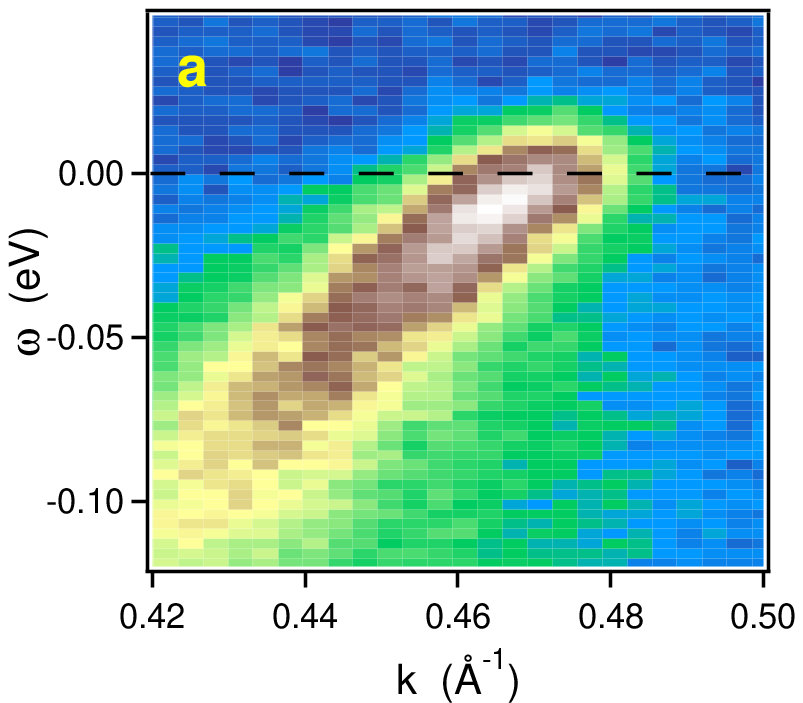}%
\includegraphics[width=4.25cm]{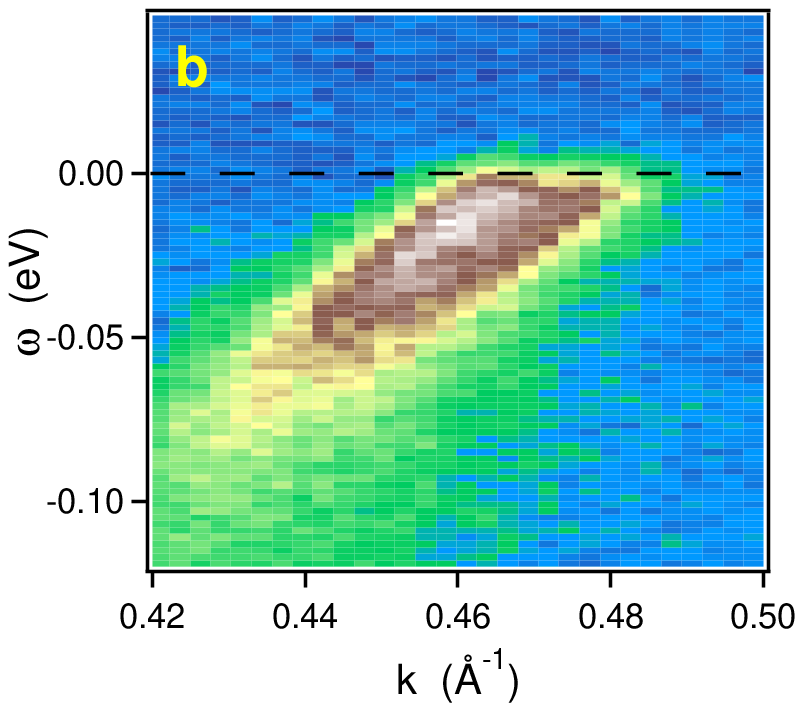}\\%
\includegraphics[width=8.4cm]{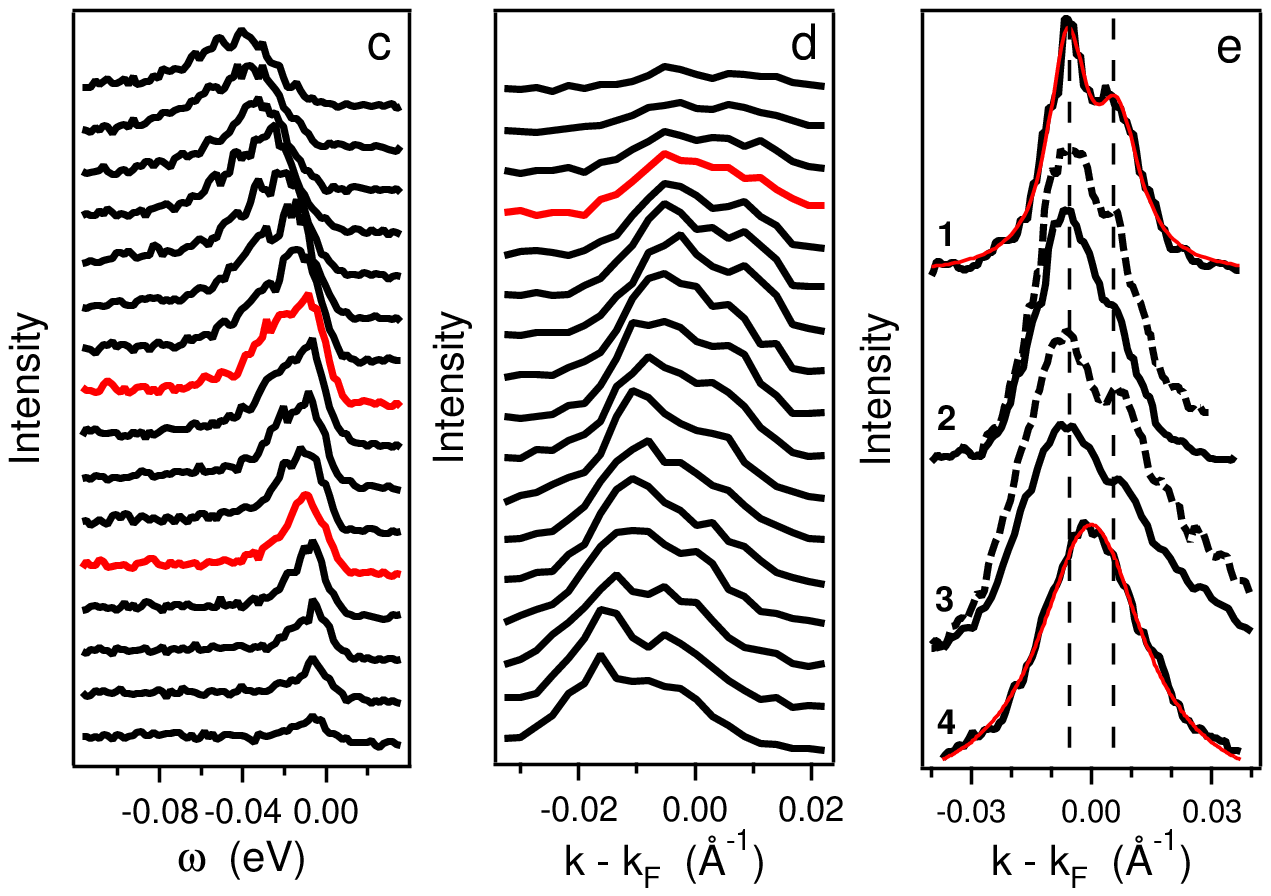}%
\caption{\label{NBS} Zoomed ARPES snapshots taken along the nodal ("$\Gamma$Y") direction of the BZ for Bi-2212 UD80 (a,b) at 27 eV and 17.5 eV excitation energy respectively \cite{NBS}. The experimental data for the same sample presented in form of energy distribution curves (EDCs) (c) and momentum distribution curves (MDCs) (d). EDCs are taken in the momentum range from $k_F -$ 0.025~\AA$^{-1}$ (top) to $k_F$ + 0.015~\AA$^{-1}$ (bottom), where $k_F$ is an average between antibonding, $k^a_F$, and bonding, $k^b_F$, Fermi level crossings; red EDCs roughly correspond to $k^a_F$ and $k^b_F$. MDCs are taken in the energy range from 3 meV (top) to $-$27 meV (bottom); $E_F$-MDC shown in red. (e) MDCs integrated in energy (along the experimental dispersion) about 10 meV from $E_F$: (1) -- Bi-2212 UD80, 1st BZ, 17.5 eV; (2) -- Bi-2212 OP89, 2nd BZ, 20 eV for the bold curve and 18 eV for the dotted curve; (3) -- Bi(Pb)-2212 OD73, 1st BZ, 17.5 eV, dashed curve results from the same MDCs but normalized to highest binding energy; (4) -- Bi-2201, 1st BZ, 17.5 eV; the red curves, when shown, represent fitting results.}
\end{figure}

A distinguishing feature of modern ARPES is the ability to resolve the bilayer splitting (BS) of the CuO conduction band in the bilayer cuprates. For the first time such a splitting has been observed for overdoped Bi-2212 \cite{FengPRL01,ChuangPRL01} and then also for optimally doped and underdoped samples \cite{ChuangXXX,KordyukPRB2002} (clearly resolved below \cite{KordyukPRB2002,BorisenkoPRB2002} and above \cite{BorisenkoPRL2003} the superconducting transition). It has been found \cite{FengPRL01,ChuangPRL01} that the observed splitting can be approximated by a momentum dependence: $t_{\perp} (\cos k_x - \cos k_y)^2 /2$, which is expected for an inter-plane hopping between two CuO layers (where $t_{\perp}$ describes the interlayer hopping mainly mediated via Cu$4s$ orbitals). The splitting along the node is expected to be not zero but much less than the maximum splitting at the saddle-point \cite{AndersenJPCS95}. In order to answer the question whether the splitting really vanishes in nodal direction, we performed precise measurements in the low excitation energy range (17--22 eV) \cite{NBS}. The total energy resolution was set to 10 meV, the angular resolution of the analyser was 0.15$^{\circ}$. To conclude on the existence of the nodal BS one should ensure taking the spectra from exactly the nodal direction. We determined the nodal direction measuring the Fermi surface (FS) maps with 0.5$^{\circ}$ step in azimuth angle (for the details about the experimental setup see \cite{BorisenkoPRB2001}). The spectra which we qualify as nodal and discuss below are taken from the FS cuts with the smallest $k_F$ which also turned out to have the steepest dispersion and the smallest leading edge gap \cite{KordyukPRB2003}.

Fig.~\ref{NBS} represents the experimental evidence for the nodal splitting. Panels a and b show ARPES snapshots taken along the nodal ($\Gamma$Y) direction of the BZ for Bi-2212 UD80 sample at 27 eV and 17.5 eV excitation energy respectively (note that the momentum-energy window is much smaller here than on standard snapshots like in Fig.~\ref{XMY}). While at 27 eV one can see only one band crossing the Fermi-level, two bands are clearly visible at 17.5 eV. One can also notice the presence of two bands in the energy distribution curves (EDCs), see panel c, extracted from the same dataset and, more explicitly, on the momentum distribution curves (MDCs) presented in panel d. In panel e (curve 1), in order to improve statistics, we integrate the MDCs along the experimental (renormalized) dispersion in the energy range 10--20 meV around $E_F$, where the MDC width does not vary dramatically. Panels a-d show one example but we observe the same effect on a number of samples of different doping level, with and without 5x1 superstructure \cite{NBS}. 

The dependence of matrix elements on excitation energy for the nodal point in the 1st BZ exhibits a local maximum at about 17.5 eV for both the total intensity from bilayer split band and the intensity from the bonding band compared to its antibonding counterpart. For the nodal point in the 2nd BZ the dependence on matrix elements is different and the bonding band is the most pronounced for $h\nu$ = 20--21 eV excitation energy: the MDCs 2 in Fig.~\ref{NBS}e show how the bonding band peak appears when going from 18 eV (dotted curve) to 20 eV (bold curve). The observed excitation energy dependence of the effect is in accord with recent calculations of ARPES matrix elements \cite{SahrakorpiPRB2003} which show that at low energy range the emissions are dominated (peaked at about 18 eV) by excitation from just the O sites.

To extract precise BS values we fit the integrated MDCs to a superposition of two independent Lorentzians (an example of a fitting curve is shown in Fig.~\ref{NBS}e). For the presented dataset the splitting in momentum $\Delta k$ = 0.012(1)~\AA$^{-1}$ which corresponds to 48(4) meV bare band splitting (for bare Fermi velocity $v_F$ = 4.0~eV\AA ~\cite{KordyukPRB2003,Bare}) or 23 meV splitting of the renormalized band (renormalized Fermi velocity $v^R_F$ = 2.0~eV\AA). For other bilayer samples the values are similar and in a good agreement with the LDA band structure calculations \cite{NBS}. This enables us to assign the splitting predominantly to vertical inter-plane hopping between O2$p_\sigma$ orbitals and to conclude on the lack of any electronic confinement to single planes within a bilayer in Bi-2212 due to strong correlations. We note that such a careful comparison between theoretical and experimental values cannot be done anywhere in the Brillouin zone except the node due to gap opening (along the nodal direction, analysing MDCs, we determine the difference in $k_F$ for these two bands while in other $k$-regions one can mainly rely on EDC analysis in which the relation of EDC peak position with the position of the band is highly model dependent, e.g. see \cite{KordyukPRL2002}). 

\section{Simple physics}

Apparently, the nodal splitting found in Bi-2212, if not taken into account, can intricately complicate the nodal spectra. Even if unresolved, it should influence the quantities derived from these spectra, e.g.~the renormalized dispersion or scattering rate. It is observed as a dependence of these quantities on excitation energy. In the following we try to eliminate the influence of the splitting choosing an appropriate excitation energy. We focus on the experimental dataset taken at 27 eV in order to find out whether the quasiparticle spectral weight from only one band can be described by "simple physics", i.e. in terms of quasiparticle self-energy \cite{AGD}.

\begin{figure}[b!]
\includegraphics[width=7.6cm]{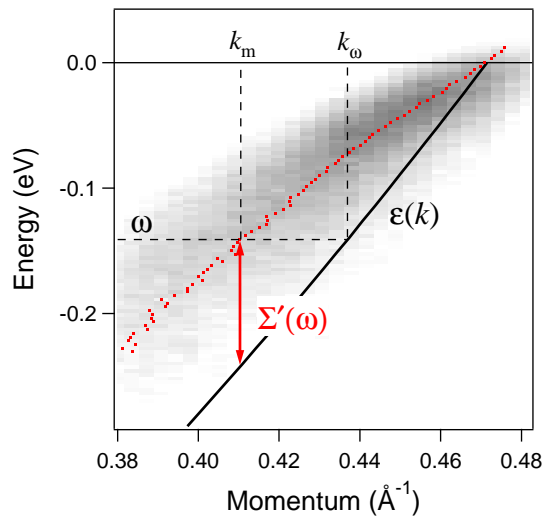}% 
\caption{\label{GX} Bare band dispersion (solid line) and renormalized dispersion (red points) on top of the spectral weight of interacting electrons. Though intended to be general, this sketch represents the nodal direction of an underdoped Bi-2212 \cite{Bare}.}
\end{figure}

Fig.~\ref{GX} illustrates the basics of the nodal spectra analysis within the self-energy approach. The black solid line represents a non-interacting case when the spectral function is a delta function with the pole $\omega - \varepsilon(\mathbf{k}_{\omega}) = 0$. $\varepsilon(\mathbf{k})$ is called "bare dispersion". A simple electronic interaction leads to shifting and broadening of the non-interacting spectral function and the resulting picture is essentially that which is measured by ARPES: the blurred region in Fig.~\ref{GX} illustrates the distribution of the quasiparticle spectral weight. Such an interaction can be described by introducing a quasiparticle self-energy $\Sigma = \Sigma' + i\Sigma''$, an analytical function the real and imaginary parts of which are related by the Kramers-Kronig (KK) transformation \cite{Landau}. Neglecting the momentum dependence of the self-energy, the MDC at certain $\omega$ exhibits a Lorentzian lineshape \cite{KaminskiPRL01} with the maximum at $k_m(\omega)$ determined by $\Sigma'(\omega) = \omega - \varepsilon(k_m)$, which is illustrated in Fig.~\ref{GX} by the double headed arrow (red squared symbols show the renormalized dispersion). In the region where the bare dispersion can be considered as linear ($\varepsilon = v_F k$), the MDC width $W$ (the half width at half maximum) is proportional to $\Sigma''$: $\Sigma''(\omega) = -v_F W(\omega)$. Thus, the determination of both the real and imaginary parts of the self-energy requires the knowledge of the bare dispersion. The KK transformation, giving an additional equation to relate these functions, opens the way to extract all desired quantities from the experiment. 

For example, one can express the coupling strength $\lambda = - (d\Sigma'(\omega)/d\omega)_{\omega=0} = v_F/v_R -1$, where $v_R$ is the renormalized Fermi velocity, as \cite{Bare}
\begin{eqnarray}\label{E3}
\lambda = \frac{-2}{\pi}\,\,PV\int_0^\infty{\frac{\Sigma''(\omega) - \Sigma''(0)}{\omega^2}\,d\omega} \equiv -\mathbf{D} \Sigma''.
\end{eqnarray}
Using the above definition of the $\mathbf{D}$ operator, $v_F^{-1} = v_R^{-1} - \mathbf{D} W$, or $1 + \lambda = 1/Z $, where
\begin{eqnarray}\label{E4}
Z = 1 - v_R \mathbf{D} W
\end{eqnarray}
is the coherence factor ($0 < Z < 1$).

In case $W(\omega)$ decays to zero or saturates on the scale covered by experiment, as it is expected for the scattering by phonons \cite{VallaPRL99}, the parameters $v_F$, $\lambda$ or $Z$ can be easily determined from the experimental values of $v_R$ and $\mathbf{D}W$. In cuprates, however, the MDC width $W$ along the nodal direction does not decrease or even saturate in the whole experimentally accessible energy region (up to $\omega_m =$ 0.5 eV) and one can make only a rough estimation expanding $\mathbf{D}W = \mathbf{D}_0^{\omega_m}W_{exp} + \mathbf{D}_{\omega_m}^\infty W_{mod}$, where $W_{exp}$ is the experimentally determined function of $\omega$, and $W_{mod}$ is a model function which depends on both the high energy cut-off, $|\omega_c| > |\omega_m|$, above which $\Sigma''(\omega)$ starts to decrease or saturate and a model for these high energy tails. For a simple estimation one can take $W_{exp} = \alpha \omega^2$ and $W_{mod} = \alpha \omega_m^2$ which gives $\mathbf{D}_0^{\omega_m}W_{exp} = \mathbf{D}_{\omega_m}^\infty W_{mod}$, demonstrating that the contribution of the tails can be essential. 

\begin{figure}[t!]
\includegraphics[width=7.2cm]{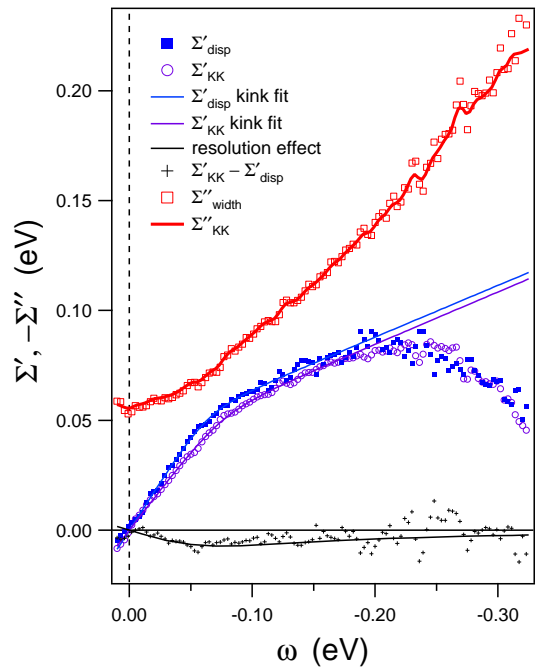}% 
\caption{\label{ImRe} Real and imaginary parts of the self-energy extracted from the experiment with the described procedure. A complete coincidence between the corresponding parts of the self-energy calculated from the two different experimental functions, the MDC dispersion and MDC width, demonstrates the full self-consistency of the ARPES data treated within the self-energy approach \cite{Bare}.}
\end{figure}

\begin{figure*}[]
\includegraphics[width=6cm]{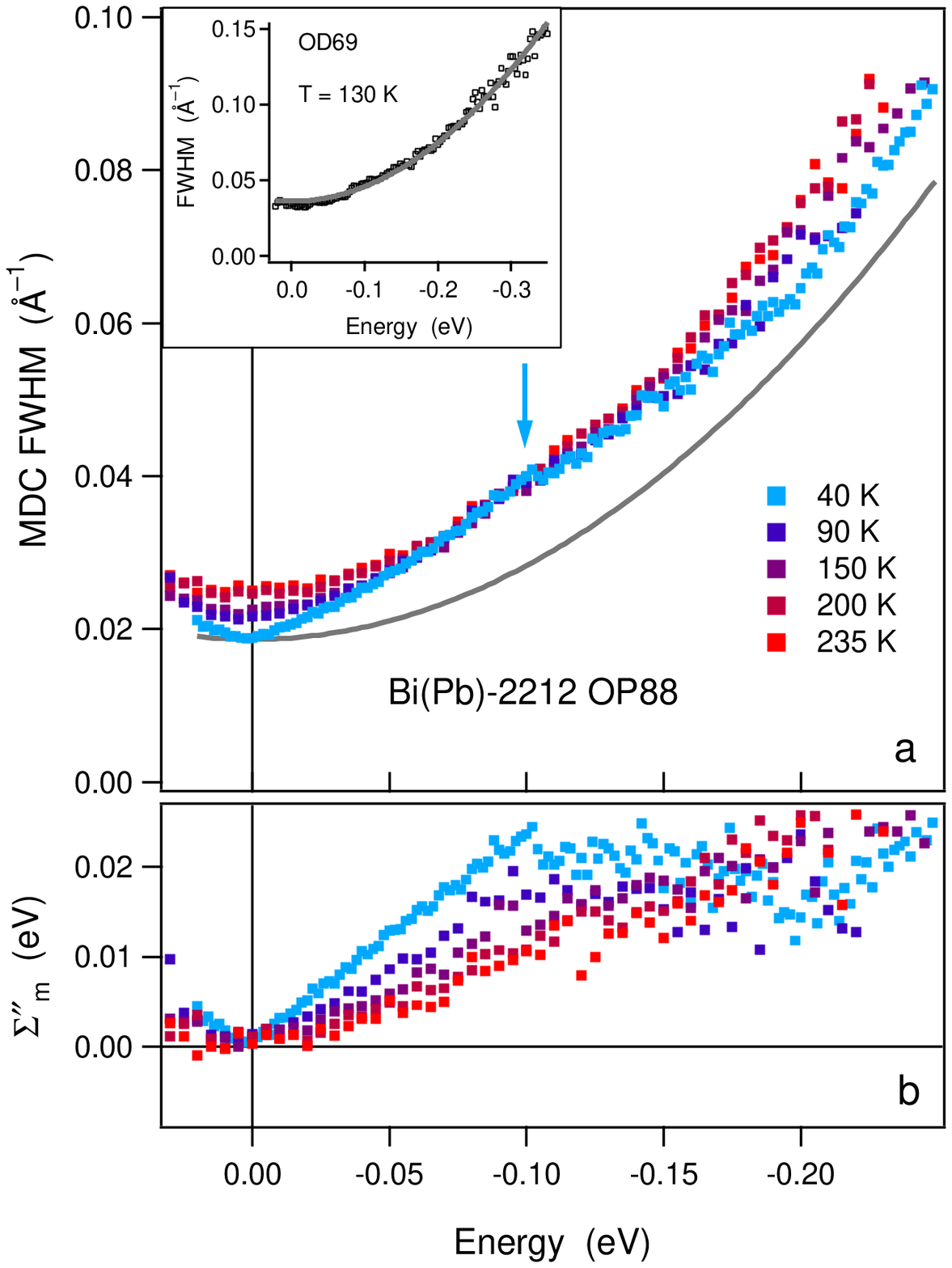}%
\includegraphics[width=6cm]{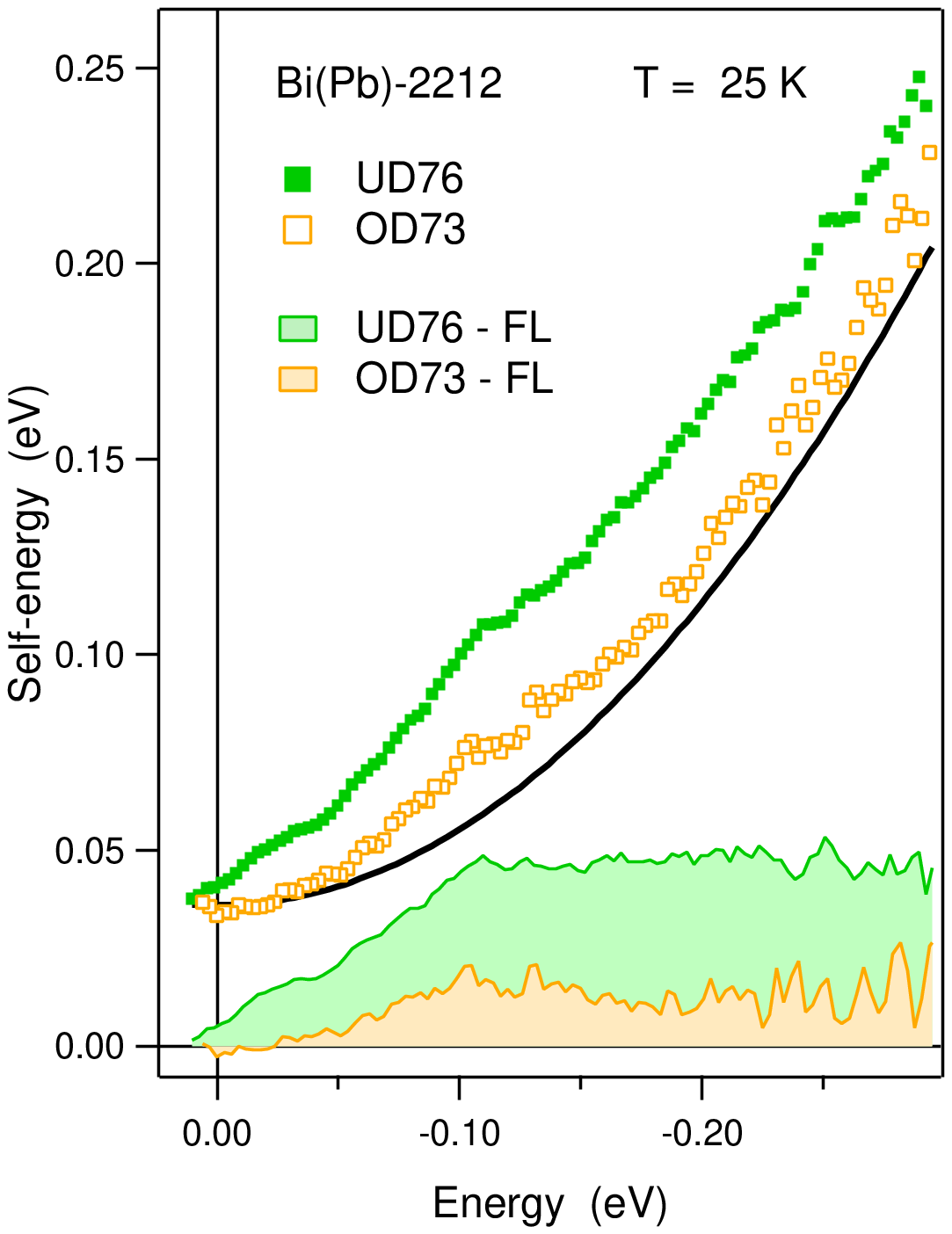}%
\caption{\label{Drop} (a) Temperature dependence of the scattering rate for the nodal quasiparticles in optimally doped Bi(Pb)-2212. The gray solid line represents a contribution from the usual Auger decay (Fermi liquid parabola) \cite{AGD} obtained by fitting the data for highly overdoped sample (OD69) at 130 K (see inset). (b) Result of a subtraction of the Fermi liquid parabola for each temperature in terms of the imaginary part of the self-energy (the FWHM/2 is multiplied to the bare Fermi velocity $v_F$ = 4 eV\AA). (c) Strengthening of the scattering mode with underdoping. Comparison of the imaginary part of the self-energy of nodal quasiparticles in Bi(Pb)-2212 underdoped ($T_c$ = 76 K) and overdoped ($T_c$ = 73 K) samples at 25 K. The shaded areas represent the contributions from the magnetic scattering obtained by subtraction of the FL parabola \cite{KordyukPRL2004}.}
\end{figure*}

Fortunately, we have found that with some assumptions and considering a sufficiently wide energy range (up to 300 meV) one can precisely determine the bare dispersion while the cut-off energy for $\Sigma''(\omega)$ and its tails remain undefined. In this procedure we use the parabolic bare dispersion which is a good approximation of the dispersion derived from the tight-binding fit of the Fermi surface \cite{KordyukPRB2003}. This assumption brings some corrections to the self-energy parts derived from the experimental data: 
\begin{eqnarray}
\label{E5}\Sigma'(\omega) = \frac{v_F}{2} (k_m^2(\omega)-k_F^2) + \omega, \\
\label{E6}\Sigma''(\omega) = -v_F W(\omega) \sqrt{k_m^2(\omega)-W^2(\omega)}.
\end{eqnarray}
The KK transform completes the system: 
\begin{eqnarray}
\label{E7}\Sigma'(\omega) = \mathbf{KK} \Sigma''(\omega).
\end{eqnarray}
Eqs. (\ref{E5}) and (\ref{E7}) give two independent ways to calculate $\Sigma'(\omega)$, and this is a core of the procedure which is described in details in \cite{Bare}. 

Fig.~\ref{ImRe} shows the results for an underdoped Bi(Pb)-2212 ($T_c =$ 77 K) at 130 K, i.e. in the pseudo-gap state. $\Sigma'(\omega)$ calculated in two ways have appeared to be identical for the certain $v_F = 3.82 \pm 0.17$ eV\AA. This gives the following interaction parameters: $\lambda = 0.87 \pm 0.12$, $Z = 0.54 \pm 0.03$. The extracted bare band dispersion is in good agreement with the band structure calculations \cite{NBS} and allows one to quantify the self-energy of the electronic excitations on the real energy scale.

The demonstrated self-consistency can be considered as a validity criterion for photoemission spectra, since it weeds out not only a "complex structure" like splitting or admixture of other bands but also artificial effects like inhomogenity of the detector, etc. Considering the intimate relation between $\Sigma'$ and $\Sigma''$ one can distinguish two energy scales in Fig.~\ref{ImRe}. One, at about 200 meV, corresponds to the maximum of $\Sigma'(\omega)$ and is related to the cut-off energy for $\Sigma''(\omega)$. Another, at 70 meV, is a famous "kink", it develops as a sharp bend between two linear segments of $\Sigma'(\omega)$ (a peak in $d^2\Sigma'(\omega)/d\omega^2$), is related to the "drop" in the scattering rate \cite{KoitzschPRB2004,ZhouNature03,KordyukPRL2004}, and commonly explained as an interaction with a bosonic mode \cite{VallaSci99,BogdanovPRL00,KaminskiPRL01}. It is important to distinguish these two scales in order to find out the nature of the coupling boson.

As far as the kink on the dispersion appears as just a sharpening of a bend of the same sign in the experimental dispersion which is present at every temperature and doping \cite{JohnsonPRL01,KoitzschPRB2004}, we focus on the "scattering rate kink" \cite{KordyukPRL2004} which is much more convenient in this sense because it develops on top of the strong normal state scattering of the opposite curvature.

Studying a number of samples of different doping level at different temperature we have found that the scattering rate kink makes it possible to distinguish between the different scattering channels \cite{KordyukPRL2004}. We argue that the main contribution to the scattering can be well understood in terms of the conventional Fermi liquid model (FL) \cite{AGD} while the additional doping dependent contribution apparently has a magnetic origin.

Fig.~\ref{Drop}a shows the scattering rate (in momentum units) as a function of frequency for optimally doped Bi(Pb)-2212 OP89 for different temperatures. A sharp kink seen in $\Sigma''(\omega)$ at 0.1 eV (indicated by the arrow) at 40 K (below $T_c$ = 88 K) gradually vanishes with increasing temperature. Another important finding is that the high binding energy tail of $\Sigma''(\omega)$ shifts upwards with temperature similar to the $\Sigma''(0)$ value. This shift, being in agreement with optical conductivity results \cite{vanderMarelNature03}, contradicts, in fact, the \textit{marginal} FL scenario \cite{VarmaPRL89}, according to which $\Sigma''(\omega, T) \propto \max(|\omega|, T)$. Such a shift of the whole curve is expected within the FL model when the scattering rate is determined by an Auger-like decay (the process where the hole decays into two holes and one electron \cite{AGD}) that gives $\Sigma'' \propto \omega^2 + (\pi T)^2$. The FL behaviour is generally expected for overdoped samples \cite{VarmaPR02}, and in Fig.~\ref{Drop}a we add the FL parabola (solid line) which perfectly fits the scattering rate for an OD69 sample above $T_c$ in the whole binding energy range. This parabola evidently describes the main contribution to $\Sigma''$ at any temperature. The additional contribution, which is seen as a hump on top of the FL parabola, must originate from an additional interaction which can be responsible for the unusual properties of the cuprates. In Fig.~\ref{Drop}b we evaluate this interaction subtracting the FL parabola for each temperature and setting the resulting offsets to zero. 

Therefore, this additional contribution decreases with increasing temperature, and we have found that it vanishes above $T_c$ for the overdoped samples, but persists at higher temperatures, presumably up to $T^*$ for optimally doped and underdoped samples \cite{KordyukPRL2004}. In Fig.~\ref{Drop}c we compare the absolute values of $\Sigma''(\omega)$ for underdoped (UD76) and overdoped (OD73) Bi(Pb)-2212 at $T$ = 25 K. The room temperature scattering rates for these two samples coincide within the experimental error bars.  It is seen that at low temperature the underdoped sample exhibits a much higher scattering rate with a more pronounced kink that has a tendency to disappear completely at higher doping levels \cite{KoitzschPRB2004}. The differences between these data and the FL parabola (solid line, the same as in Fig.~1a) demonstrate that the additional scattering channel of the nodal quasiparticles is highly doping dependent which is difficult to reconcile with the phonon scenario \cite{ZhouNature03,LanzaraNature01}, leaving space for magnetic excitations as the only bosons responsible for this additional channel \cite{FongPRB00,EschrigPRB03}.  

Although our findings support the magnetic nature of the doping dependent channel in the scattering rate, its origin is still to be understood. There are two main suspects for the scattering kink and dispersion kink (in terms of a sharp feature that was discussed above): the magnetic resonance and gapped spin-fluctuation continuum \cite{ChubukovKink}. As far as the magnetic resonance at ($\pi,\pi$) is believed to be sharp in energy and momentum it seems unlikely to connect the nodal region to another part of the FS by such a scattering. This increases the possibility to describe the kinks by the gapped spin-fluctuation continuum alone and it has been recently shown \cite{ChubukovKink} that the gapped magnetic spectrum can well describe the dispersion kink feature. Nevertheless, we believe that the magnetic resonance cannot be ruled out because of the presence of the so called "shadow band" \cite{Aebi} which, as we have recently shown \cite{KoitzschSB}, is not a diffraction replica, but a real band of the CuO layers, and consequently provides an easy way for the quasiparticle from the main band to scatter. In order to find out which mechanism is dominant, more accurate and systematic investigations of both kink features are needed.

\section{Conclusions}

We performed a detailed investigation of the nodal direction of Bi-2212 in wide ranges of doping, temperature and excitation energy. Even along this direction, the electronic band has appeared to be split, but choosing an appropriate excitation energy we have succeeded to single out the quasiparticle spectral weight distribution from only one bilayer split band and show that it can be self-consistently described within the quasiparticle self-energy approach. Focusing on the scattering rate as a function of binding energy, temperature and doping we have distinguished two main channels in the electron scattering. While the main doping independent channel can be well understood in terms of the conventional Fermi liquid model, the additional doping dependent channel implies a magnetic origin. Our findings prove the existence of well defined quasiparticles even for the underdoped Bi-2212 ($T_c$ = 77 K) in the pseudogap state.

We acknowledge the stimulating discussions with A. Chubukov, I. Eremin, A. N. Yaresko, S.-L. Drechsler, B. Keimer and Yoichi Ando, technical support by R. Follath, O. Rader, and R. H\"ubel. The project is part of the Forschergruppe FOR538 and is supported by the DFG under grants number KN393/4 and 436UKR17/10/04. H.B. is grateful to the Swiss National Science Foundation and its NCCR Network "Materials with Novel Electronic Properties" for support.

\end{document}